# ACCESSIBILITY EVALUATION OF MAJOR ASSISTIVE MOBILE APPLICATIONS AVAILABLE FOR THE VISUALLY IMPAIRED

Saidarshan Bhagat[1], Padmaja Joshi[2], Avinash Agarwal[3], Shubhanshu Gupta[4]

[1] (saidarshan@cdac.in), C-DAC Mumbai, India [2] (padmaja@cdac.in), C-DAC Mumbai, India [3] (avinash.70@gov.in), Telecommunications Engineering Centre, New Delhi, India [4] (shubhanshug@cdac.in), C-DAC Pune, India

**Abstract** – People with visual impairments face numerous challenges in their daily lives, including mobility, access to information, independent living, and employment. Artificial Intelligence (AI) with Computer Vision (CV) has the potential to improve their daily lives, provide them with necessary independence, and it will also spawn new opportunities in education and employment. However, while many such AI/CV-based mobile applications are now available, these apps are still not the preferred choice amongst visually impaired persons and are generally limited to advanced users only, due to certain limitations. This study evaluates the challenges faced by visually impaired persons when using AI/CV-based mobile apps. Four popular AI/CV-based apps, namely Seeing AI, Supersense, Envision and Lookout, are assessed by blind and low-vision users. Hence these mobile applications are evaluated on a set of parameters, including generic parameters based on the Web Content Accessibility Guidelines (WCAG) and specific parameters related to mobile app testing. The evaluation not only focused on the guidelines but also on the feedback that was gathered from these users on parameters covering the apps' accuracy, response time, reliability, accessibility, privacy, energy efficiency and usability. The paper also identifies the areas of improvement in the development and innovation of these assistive apps. This work will help developers create better accessible AI-based apps for the visually impaired.

## 1. INTRODUCTION

Vision is considered the most important of the five basic human senses [1]. It plays a vital role in our daily activities, including reading, learning, driving, watching TV, and even recognizing familiar faces. Persons with blindness or low-vision, referred to as Visually Impaired Persons (VIPs) in this paper, face a multitude of challenges in their daily lives and have to largely depend on family and friends for performing basic tasks. Challenges include reading prints and emails, recognising currencies, identifying medicines, and even identifying the colour of their clothes [2]. They also face hardships in activity and participation [3], using smartphones [4], using public transportation [5], and many more. Visually impaired children are likely to have poorer educational outcomes [6], thus affecting overall development and employability.

The World Report on Vision released by World Health Organization (WHO) in 2019 [7] mentions that vision impairment is widespread and estimates that globally at least 2.2 billion people have vision impairment. The proportion of VIPs is greater among low and middle-income countries, older people, and rural communities [7]. In India, 2.21% of the population has one or more disabilities, with 19% of them having visual disabilities [8], [9]. To focus global attention on vision impairment, including blindness, the World Health Organisation organises World Sight Day on the second Thursday of October every year [10].

Enhancing accessibility to the visually impaired will play a vital role in attaining various Sustainable Development Goals (SDGs), including reducing poverty, improving health and education, increasing work efficiency, and promoting equity [6]. The United Nations recognises the need to ensure access to affordable and accessible assistive technology to achieve the Sustainable Development Goals [11].

Currently, we are undergoing a major technological transformation, and Artificial Intelligence (AI) is at the centre of many of these innovations. Over the past decade, we have heard many stories about AI and its fascinating applications. But today, we are seeing AI solutions in action. From Computer Vision (CV) to speech recognition, AI is the backbone of these emerging technologies.

One of the most noteworthy target users of AI is people with disabilities. To serve this significant section of society, innovative AI and CV-based solutions are being developed for people with visual disabilities to improve their daily lives. CV assists in everyday tasks, such as navigation, identifying objects, recognising persons, reading, retrieving information, and entertainment [12]. These AI/CV-based applications are true game changers, as they allow people with disabilities to live with greater independence and enjoy immense liberty [13], [14].

VIPs have increasingly opted for smartphone-based apps as affordable assistive devices due to their portability, affordability, and ease of use [15], [16]. Smartphones are widely available and serve as multifunctional mobile devices. They come with high-end processors, cameras, microphones, speakers, and various sensors such as GPS, accelerometer, gyroscope, pedometer and compass. They can run data processing apps and exchange data wirelessly with external servers and cloud platforms [24], [26]. Further, the same smartphone can install multiple assistive apps with different focus areas.

Currently, there are many AI/CV-based assistive mobile applications for the visually impaired, but some of these solutions struggle with accuracy and response time. However, despite the life-changing features that these apps offer to VIPs, their low download numbers on app stores indicate that they are not yet popular among the target users. While some advanced users are using these apps, a large proportion of visually impaired individuals still do not find them suitable for their needs.

This paper focuses on examining the gap between the requirements of VIPs and the performance of AI/CV-based mobile applications. The aim is to identify functional shortcomings and technical challenges that VIPs face, including accuracy, response time, reliability, accessibility, privacy, usability, energy efficiency, and safety aspects. An online survey was conducted among blind and low-vision individuals to gather information on the AI-based apps they use, why they use them, the challenges they face when using them, and their expectations. Through this survey and market research, we identified four popular mobile applications for evaluation, which are the focus of this paper. Furthermore, several blind and low-vision individuals evaluated these apps based on a set of parameters, including generic parameters based on the Web Content Accessibility Guidelines (WCAG) [20] and specific parameters related to mobile app testing.

The remaining paper is organised as follows: Section 2 provides a review of the recent research on the topic. In Section 3, the selected applications are described along with the evaluation parameters and methodology used. Section 4 presents observations of this audit. In Section 5, we discuss the results and challenges VIPs face when using AI/CV-based accessibility solutions. Section 6 proposes technical and functional improvements to make these applications more usable, accessible and inclusive. The paper concludes with remarks and outlines the scope for further work in this field in Section 7.

## 2. LITERATURE REVIEW

Many assistive mobile apps for the visually impaired are available on both Android and iOS platforms [18]. Their focus areas include screen reading and writing (e.g., TalkBack, KNFB Reader, BrailleTouch, Voice Dream), navigation (e.g., Nearby Explorer, Ariadne G.P.S., BlindSquare), colour identification (e.g., Color Picker ColorID Free), object identification (e.g., ScanLife Barcode, QR Reader, LookTel Money Reader), as well as news apps (e.g., NFB-Newsline, Blind Bargains, AccessWorld) [16], [19].

We examined the existing research that compares different assistive applications for visually impaired individuals. However, there is a limited amount of literature on this topic, with only a few comparative studies available. Granquist et al. [23] conducted a study comparing Orcam MyEye 1, a portable device, and Seeing AI, an iPhone and iPad application. This study focused solely on the print-to-speech or text recognition properties of these two AI vision aids and relied on feedback from seven visually impaired participants. Similarly, Neat et al. [24] evaluated three CV-based apps (Seeing AI, Spot+OCR, and Guided OCR) based on their optical character recognition abilities, using a quantitative metric called the True Positive Ratio (TPR). This study also relied on feedback from seven blind participants and only examined the text recognition capabilities of these apps. In another study, Gil et al. [25] assessed the usability and accessibility of two mobile applications, LookOut (Android) and Seeing AI (IOS), for reading text documents, recognising colours, identifying people, and recognising paper money as a support tool for visually impaired individuals. This study used a qualitative questionnaire-based survey to gather feedback. COA, Dominique M. Dockery et al. [26] conducts a phone survey with visually impaired individuals to discover that Seeing AI and Be My Eyes are the most commonly use apps to help low-vision patients with activities of daily living.

Hélène Walle et al. [27] analyses some existing surveys and studies some current work based on acquired data, learned models, and human–computer interfaces. Apart from the research in AI-based applications, there is some research going in the area of AI for accessibility and assistive technology. Sayeda Farzana Aktar [28] focuses on utilizing AI/ML algorithms to enhance accessibility for People with Disabilities (PwDs) in three areas: public buildings, homes, and medical devices. Budrionis et al. [29] reviewed smartphone-based CV travelling aids for blind and visually impaired individuals to offer an overview of recent electronic travel aid research prototypes that employ smartphones for assisted orientation and navigation in indoor and outdoor spaces by providing additional information about the surrounding objects. Nilanjan Chakraborty et al. [30] discusses how artificial intelligence has helped in improving the living standards of persons with disabilities, its future possibilities, and potential legal issues concerning the implication of artificial intelligence in India.

The focus of the evaluation in this paper is different from earlier studies and reviews. This study evaluates AI-CV-based mobile apps in a partially qualitative and quantitative manner. The evaluation also considers the feedback and usage experience from persons with disabilities. Along with the AI and CV-based features we have also evaluated the usability, compatibility and privacy aspects of these apps.

## 3. METHODOLOGY

This section explains the methodology used to select mobile apps for the study, outlines the parameters chosen for evaluation and describes the process followed for evaluating the apps.

### 3.1 Methodology of app and evaluation parameter selection

Abundant AI+CV-based mobile apps are available on official application stores for Android and iOS devices. Evaluating all may not be possible, but as the aim is to see the impact on a larger audience, the following factors were focused while selecting the apps for evaluation.

We conducted an online survey on a highly popular mailing list that caters to individuals with disabilities. This survey was specifically targeted at blind and low vision users, aiming to gain insights into their preferred AI and computer vision-based recognition applications available on both the leading mobile platforms Android and iOS. During this survey, we encouraged end users to share their favourite features from these applications and to provide feedback regarding the challenges they encountered across various parameters. Furthermore, we collected their valuable suggestions for potential enhancements in future iterations of these applications. The data gathered from this survey played a pivotal role in our process, aiding us in selecting the apps for evaluation and, of equal importance, to identify four popular mobile applications and determining the key parameters for our evaluation such as accuracy, performance, reliability, accessibility, privacy, security, compatibility, energy efficiency, and usability.

The target user group for this evaluation is from developing countries and many individuals may not be able to afford paid apps due to economic conditions, hence affordability is the first factor. Hence the chosen apps are free of charge or without a subscription but provide essential features.

As the focus is on the apps having a larger impact on visually challenged people, the factors chosen are the availability of the apps across widely used platforms covering 99% of the global market which are Android or iOS. Secondly, the popularity of the apps. The number of downloads, user ratings and user reviews are the parameters used to identify the popularity of the app. The selected apps have a minimum of 10 000 downloads, user ratings of at least 3.5 on app stores, and at least 150 user reviews.

### 3.1.1 Microsoft Seeing AI

Seeing AI [31] by Microsoft is exclusively available for the iOS platform. It is one of the most popular AI/CV-based apps available around the globe. It facilitates multiple tasks to provide description of camera captured images such as short text, documents, handwritten text, product barcodes, persons, scene, currency, light brightness, perceived colour etc.

### 3.1.2 Envision

The Envision [32] app is available on both the Android and iOS platforms. Versions on each platform differ in their features and overall results. It describes surroundings, describe scenes, detect colours, find objects, scan product barcodes, finds nearby persons with pretrained faces, has smart guidance for text on the fly, reads out handwritten and printed text, auto-detects language and text layout, has the facility to export detected text and has batch scan text.

### 3.1.3 Supersense

Supersense [33] is a market leader mobile scanner application to scan text, money and objects for blind and visually impaired users. It is equipped with novel computer vision and machine learning models and is available on both the iOS and Android platforms: however, on the iOS platform, most of the important features are available exclusively to paid premium users.

### 3.1.4 Google lookout

The Google lookout [34] app helps automatically read and scan text, recognise products and describe objects to help blind and visually impaired people learn about their surroundings. This app is exclusively available on the Android platform.

**Table 1** – Mobile app selection factors

| Application name | Microsoft Seeing AI | Super-sense AI for blind | Super-sense AI for blind | En-vision AI | En-vision AI | Look-out – Assisted Vision |
|---|---|---|---|---|---|---|
| Platform | iOS, iPadOS | Android | iOS | Android | iOS | Android |
| Release date (main) | Jul 2017 | Feb 2019 | Mar 2020 | 2017 | -- | Mar 2019 |
| Downloads | 20M | 50K | -- | 100K+ | -- | 100K+ |
| Size | 302.9 MB | 104 MB | 307.9 MB | 54 MB | 252.4 MB | 26.63 MB |
| Rating | 4.4 | 3.9 | 4.3 | 4.6 | 4.2 | 4.1 |
| No. of reviews | 497 | 930 | 207 | 4.3K | 325 | 2.4K |

## 3.2 Evaluation parameters

The first focus was identifying the apps for evaluation and then identifying the parameters of evaluation to meet our objective. This was achieved through the online survey conducted with members of the blind and low-vision community. The identified parameters should be independent of the platform Android or iOS. The key parameters focused for the evaluation are accuracy, performance, reliability, accessibility, privacy, security, compatibility, energy efficiency, and usability.

### 3.2.1 Accuracy

Accuracy is crucial for deciding the application's utility. Accuracy is dependent on four essential aspects in this case which are: printed text recognition, handwritten text recognition, scene/object recognition and recognition of currency notes of different denominations. The accuracy is then classified as good, average and poor.

### 3.2.2 Performance

Performance refers to the time taken by the application to respond. We evaluated the response time taken by the selected apps to complete the recognition process for text, objects and currency notes using both quantitative and qualitative methods.

### 3.2.3 Reliability

Reliability is an application's ability to generate consistent results and is crucial for people with disabilities who use these AI/CV-based apps for daily tasks. We assessed this parameter by performing multiple recognitions of similar text, objects, and currency notes and measuring the number of errors.

### 3.2.4 Accessibility

Accessibility is a decisive parameter that evaluates the access of the app features to visually disabled users based on the popular WCAG principles - Perceivable, Operable, Understandable and Robust (POUR) [20]. We assessed accessibility qualitatively by testing the app's user interface using various assistive technologies, with poor labelling, poor colour contrast, and a non-customisable interface considered anomalies in app design.

### 3.2.5 Privacy and security

Privacy and security become a high priority for AI/CV-based applications as they control sensitive user data. We evaluated the selected apps against this qualitative parameter by reviewing their privacy and data management policies. We also checked the user's control to delete or verify the data.

### 3.2.6 Compatibility

Compatibility parameters check the compatibility level of the app with various assistive technologies such as screen readers and magnifiers. We assessed these parameters of the selected apps qualitatively.

### *3.2.7 Energy efficiency*

Energy efficiency is a crucial parameter when it comes to the daily usage of any smartphone. We evaluated the battery performance of our devices while using the selected apps on Android and iOS. Apps that drastically reduce the battery are rated poor in energy efficiency and tend to be avoided by users.

### *3.2.8 Usability*

Usability is a parameter that assesses the user-friendliness of an application by evaluating the user interface and navigational controls. We tested the ease of operation and overall usability of the application, checked its language offerings, availability of help modules, and options to switch UI themes, and evaluated it through qualitative methods.

## 3.3 Evaluation process followed

Blind and low-vision individuals exclusively evaluated the apps according to the parameters and subparameters outlined in Section 3.2. They conducted a series of manual tests to analyse recognition results using predefined inputs such as printed paragraphs, handwritten paragraphs, currency notes and an image with natural surrounding detailed later in this section. We gathered real-time data and recorded the usage experience of the target users.

We installed the selected apps on an iPhone 13 running iOS 16 and an Android phone running Android 12. We granted the necessary permissions as requested by these applications.

For text recognition, we used one printed text and one handwritten text in different lighting conditions. OCR of mobile camera captured images under variable lighting conditions have been studied by Pushpinder Singh et al. [35]. Lighting conditions have a high impact on the accuracy of text acquisition and recognition performed using a mobile camera. For the image and scene recognition feature, we used one common image and natural surroundings. Additionally, we evaluated the Indian currency recognition feature using six different denominations in varying lighting conditions. We performed each test procedure multiple times to calculate the average response time and measure app reliability.

We used the built-in screen reader and magnifier to verify the accessibility and compatibility of all evaluations. We assessed each user interface component against the accessibility guidelines. We also reviewed the privacy and data protection policy of each application.

Lastly, we checked the battery usage of all the apps to evaluate their energy efficiency.

## 4. OBSERVATIONS AND FINDINGS

### 4.1 Accuracy

**Table 2** – Accuracy ratio in printed medium paragraphs

| Sl. No. | Lighting condition | Seeing AI for iOS | Super-sense AI for Blind for Android | Super-sense AI for Blind for iOS | En-vision for Android | En-vision for iOS | Look-out for An-droid |
|---|---|---|---|---|---|---|---|
| i). | Good | 100% | 95% | 98% | 100% | 99% | 100% |
| ii). | Average | 100% | 93% | 97% | 97% | 98% | 95% |
| iii). | Low | 99% | 91% | 96% | 98% | 98% | 93% |

**Table 3** – Currency identification ratio for six denominations – rs. 10, 20, 50, 100, 200, 500

| Sl. No. | Lighting condition | Seeing AI for iOS | Super-sense AI for Blind for Android | Super-sense AI for Blind for iOS | En-vision for Android | En-vision for iOS | Look-out for An-droid |
|---|---|---|---|---|---|---|---|
| i). | Good | 100% | Not supported | Not supported | Not supported | Not supported | 100% |
| ii). | Average | 99% | Not supported | Not supported | Not supported | Not supported | 100% |
| iii). | Low | 99% | Not supported | Not supported | Not supported | Not supported | 100% |

**Table 4** – Accuracy ratio in two handwritten medium paragraphs

| Sl. No. | Lighting condition | Seeing AI for iOS | Super-sense AI for Blind for Android | Super-sense AI for Blind for iOS | En-vision for Android | En-vision for iOS | Look-out for An-droid |
|---|---|---|---|---|---|---|---|
| i). | Good | 90% | Not supported | Not supported | Not supported | Not supported | 90% |
| ii). | Average | 75% | Not supported | Not supported | Not supported | Not supported | 90% |
| iii). | Low | 75% | Not supported | Not supported | Not supported | Not supported | 75% |

## 4.2 Performance

**Table 5** – Response time in seconds

| Sl. No. | Parameter | Seeing AI for iOS | Supersense AI for Blind for Android | Supersense AI for Blind for iOS | Envision for Android | Envision for iOS | Lookout for Android |
|---|---|---|---|---|---|---|---|
| i). | Text recognition | 90% | 4-5 seconds | 3-4 seconds | 5-6 seconds | 2-3 seconds | 4-5 seconds |
| ii). | Scene/object recognition | 75% | 2-3 seconds | Not supported | Not supported | 3-4 seconds | 10-12 seconds |
| iii). | Currency recognition | 75% | 2-3 seconds | Not supported | Not supported | Not supported | Not supported |

## 4.3 Reliability

**Table 6** – Accuracy ratio in repeat tests

| Sl. No. | Parameter | Seeing AI for iOS | Supersense AI for Blind for Android | Supersense AI for Blind for iOS | Envision for Android | Envision for iOS | Lookout for Android |
|---|---|---|---|---|---|---|---|
| i). | Text recognition | 99% | 97% | 97% | 98% | 98% | 99% |
| ii). | Scene/object recognition | 100% | Not supported | Not supported | 97% | 97% | 99% |
| iii). | Currency recognition | 99% | Not supported | Not supported | Not supported | Not supported | 100% |

## 4.4 Accessibility

**Table 7** – Accessibility aspects

| Sl. No. | Parameter | Seeing AI for iOS | Supersense AI for Blind for Android | Supersense AI for Blind for iOS | Envision for Android | Envision for iOS | Lookout for Android |
|---|---|---|---|---|---|---|---|
| i). | Unlabelled controls | None | None | None | None | None | None |
| ii). | Inoperable controls | None | None | None | None | None | None |
| iii). | Controls with poor colour contrast ratio | None | None | None | None | None | None |

## 4.5 Privacy and security

**Table 8** – Privacy and security

| Sl. No. | Parameter | Seeing AI for iOS | Supersense AI for Blind for Android | Supersense AI for Blind for iOS | Envision for Android | Envision for iOS | Lookout for Android |
|---|---|---|---|---|---|---|---|
| i). | Privacy statement provided? | Yes | Yes | Yes | Yes | Yes | Yes |
| ii). | Option to delete the data from servers? | No | No | No | No | No | No |
| iii). | Prompts before taking important device permissions? | Yes | Yes | Yes | Yes | Yes | Yes |

## 4.6 Compatibility

**Table 9** – Compatibility issues

| Sl. No. | Parameter | Seeing AI for iOS | Supersense AI for Blind for Android | Supersense AI for Blind for iOS | Envision for Android | Envision for iOS | Lookout for Android |
|---|---|---|---|---|---|---|---|
| i). | Issue/glitch with screen reader | None | None | None | None | None | None |
| ii). | Issue/glitch with screen magnifier | None | None | None | None | None | None |
| iii). | Other platform issues | None | None | None | None | None | None |

## 4.7 Energy efficiency

**Table 10** – Energy efficiency

| Sl. No. | Parameter | Seeing AI for iOS | Supersense AI for Blind for Android | Supersense AI for Blind for iOS | Envision for Android | Envision for iOS | Lookout for Android |
|---|---|---|---|---|---|---|---|
| i). | Battery drain of the device during apps usage | Average | Average | Average | Average | Average | Average |
| ii). | Energy-efficiency indicator | Medium | Medium | Medium | Medium | Medium | Medium |

## 4.8 Usability

**Table 11** – Usability

| Sl. No. | Parameter | Seeing AI for iOS | Supersense AI for Blind for Android | Supersense AI for Blind for iOS | Envision for Android | Envision for iOS | Lookout for Android |
|---|---|---|---|---|---|---|---|
| i). | User friendliness of UI. | Good | Average | Average | Medium | Medium | Good |
| ii). | Is app available in multiple languages? | Yes | Yes | Yes | Yes | Yes | Yes |
| iii). | Is text recognition available in any Indian language | No | No | No | No | No | Yes |
| iv). | Is there any help material text or video available for apps features? | Yes | Yes | Yes | Yes | Yes | Yes |
| v). | Is there any option to switch UI theme? For e.g. Dark mode | No | No | No | No | Yes | No |

### 4.8.1 Microsoft Seeing AI

In our evaluation, we observed that Seeing AI is one of the most accessible and usable apps on the market. On the accuracy parameter, it performed extremely well and produced the most accurate text recognition even in average lighting conditions. It generated a few recognition errors in low-light conditions; refer Table 2 for evaluation results. Seeing AI is capable of guiding blind and low-vision users in capturing all edges of the document for better results. Scene and object recognition also worked extremely well during the evaluation. The app provides detailed feedback about the surroundings including object colour, a person's emotion and age. An important feature of currency recognition also worked well in good lighting conditions; however, it generated some inaccurate results, especially while detecting new Indian denominations. In our observations, Seeing AI stood very well on the parameter of reliability as observed in Table 6, and it was always very efficient in text, currency and scene recognition. Seeing AI has nice video tutorials for app usage and it even has a detailed privacy policy to handle sensitive user data.

An average battery drain along with some hitting issues on the test device was observed while using this app.

### 4.8.2 Envision

During the evaluation, we observed that for the iOS version, English text recognition accuracy is decent in good lighting conditions, and it generates a few recognition errors in average lighting conditions. Text recognition for Hindi and other regional languages is not provided. For the Android version, the accuracy of English text recognition has some issues, Hindi text recognition is provided but accuracy is moderate. The iOS version did not produce any recognition of handwritten text while the Android version recognised handwritten text with some errors detailed in Table 4. On the parameter of scene and object recognition, both the Android and iOS versions provided a very minimal description. On the iOS app, scene recognition took an extremely long time. Currency recognition is not provided in both versions. Both versions of this app are accessible; however, with a lot of options and features, the user interface is a little cluttered. During the usage, we did observe some battery drain on both the iOS and Android versions.

### 4.8.3 Supersense

On the iOS version, we tried scanning a page in low lighting using instant text recognition; however, it produced very inaccurate results. In good lighting conditions, results were better; refer to Table 4 for details. Similar levels of accuracy were observed in text recognition on the Android platform. Handwritten and regional text recognition is not available in either version. Scene and object recognition were not available in the free version of iOS and provided average accuracy with some wrong results in the Android version. A detailed help tutorial is provided in both the Android and iOS versions.

### 4.8.4 Google lookout

During the evaluation, it was observed that it is an efficient and fast application for blind or low-vision users. Text recognition in English, as well as Hindi, generated good results with high accuracy. The app was successful in maintaining decent accuracy even in average and low-light conditions. In terms of handwritten text recognition, the accuracy is not as good as that of printed text. It recognised Hindi handwritten text also with minimal errors, as observed in Table 4. The application automatically guides its users to manage the camera focus to capture all corners of the document. Currency recognition produced good results even in low-lighting conditions, as mentioned in Table 3. The placement and condition of the denomination did not impact the accuracy of the recognition. It even comfortably detected the new Indian denominations.

### 4.9 Overall ratings

This parameter-centric evaluation performed on the stated apps by users with disabilities shows that each application has its advantages and shortcomings. By studying the usage pattern and specific operational needs, each application is rated on a scale of high, medium and low. The most accurate, efficient, accessible, and usable application is rated as 'High', and the less user-friendly and less accurate application is tagged as 'Medium' or 'Low'. Table 12 has this overall application rating derived from the detailed quantitative and qualitative evaluation results presented in tables 2-10. This overall rating is specific to the version which we evaluated during our study, updating the quantitative data and rating as per the latest version is out of our scope.

**Table 12** – Overall ratings

| Application Name | Android | iOS |
|---|---|---|
| Microsoft Seeing AI. | Not Applicable | High |
| Supersense | Medium | Low |
| Envision | Medium | Medium |
| Lookout | Not Applicable | High |

## 5. CHALLENGES IN DEVELOPING ASSISTIVE APPS

After studying the evaluation data presented in Section 4 in the tables (Table 2(accuracy ratio in printed medium paragraphs), Table 3 (currency identification ratio for six denominations – rs. 10, 20, 50, 100, 200, 500), Table 4 (accuracy ratio in two handwritten medium paragraphs), Table 5 (response time in seconds), Table 6 (accuracy ratio in repeat tests), Table 7 (accessibility aspects), Table 8 (privacy and security), Table 9 (compatibility issues), Table 10 (energy efficiency) and Table 11 (usability)), a few challenging areas in present AI/CV-based mobile applications are observed. These aspects not only make them intricate, but also keep them away from users with low technical literacy and budget smartphone users. Analysing the challenges mentioned in this section is enormously vital in improving the upcoming AI/CV-based assistive app development and also advantageous for future innovation in this area.

### 5.1 Accuracy in different lighting conditions

This is one of the crucial issues in the solutions available presently on the market. We have experienced this at large, while detecting document images captured in low-light conditions. This leads to inaccurate recognition results and non-reliable solutions. The most significant reason behind this issue is poor camera sensors and inadequate image processing software engines.

### 5.2 Performance optimization

One of the key challenges observed in our evaluation is the performance of the application while generating results. Identifying complex images and recognising large text in a variety of font styles takes a large toll on the overall performance and overall response of the apps. Limited datasets, insufficient processing power, and most importantly cloud-based data processing are a few important reasons for this issue.

### 5.3 Complex user interface

In the race of offering a variety of features, many application developers end up cluttering the app interface and making it complicated to use. Creating nested options and less user-friendly descriptions leads to a poor and unfriendly user interface.

### 5.4 Unavailability of regional languages

This is one of the important factors while making the solutions available in developing markets like India. Out of six evaluated applications, only one has support for Hindi text recognition. The main cause is the unavailability of the required Optical Character Recognition (OCR) engine. Due to this aspect, these AI and CV-based applications are restricted to generally English speaking users and not used by people who speak regional languages.

### 5.5 Low energy efficiency

Most of the applications we evaluated had the issue of average battery drain while using the app. Due to this, the apps became low energy efficient, which makes it difficult for a user to use them on the go. Complex processing, poorly tuned software engines, heavy usage of camera sensors, and frequent cloud interaction (network I/O) may be the reasons behind this issue.

## 6. FUNCTIONAL AND TECHNICAL IMPROVEMENTS IN FUTURE WORK

Challenges discussed in the previous section need to be addressed as the highest priority to improve the overall design of AI/CV-based mobile applications. Right from accuracy to energy efficiency, everything should be fundamental in future developments. Suggestions proposed in this section will be helpful in creating the roadmap for reliable, efficient and inclusive mobile applications.

### 6.1 Local datasets

A rich dataset is key to developing highly reliable and accurate AI/CV-based solutions. As per the evaluation in Section 4, data processing on remote cloud servers is one of the main reasons for low performance. Having local datasets in a compressed format stored on the device will surely help for quick comparisons, without communicating to remote servers; this will improve the overall performance and reliability of these applications.

### 6.2 Intelligent and compact image processing engine

A smart and highly intelligent software image processing engine is key for any good AI/CV-based solution. If image processing is executed with less time and fewer errors, then output recognition is of the highest accuracy and efficiency. Software should be smart enough to remove any noise and other bad elements from the image automatically without tampering with the original content. Images captured in low-light conditions should also be processed with a highly accurate algorithm to provide the best accuracy. Software engines should be optimised for performance and energy efficiency.

### 6.3 Universally designed user interface

The user interface of these applications should be designed by adhering to the universal design principles for better usability and accessibility. Designing an efficient and easy-to-use interface should be a top priority. The application should provide a variety of interface themes for greater user-level customization. It is observed that applications designed as per the universal design principles are inclusive, accessible and provide a number of benefits to its end users [36]. While following these principles, designers also have to adhere to the specific guidelines or standards in the area, for example, the Web Content Accessibility Guidelines (WCAG) for websites and mobile applications. Appropriate automated and manual testing procedures play a significant role in achieving the maximum compliance with these standards. A number of automated tools are available to test WCAG compliance, each having their advantages and shortcomings as observed in the paper 'Comparing ten WCAG Tools for accessibility evaluation of websites' [37] by S. Kumar et al.

### 6.4 Strong data privacy policy

This is one of the important points for discussion when it comes to any AI/CV-based application. Since these apps have direct access to sensitive user data, it is very important to understand how they deal with this. A strong and very comprehensive data and privacy policy will be a must in future applications. This will help us in creating a reliable high-tech app ecosystem.

### 6.5 Multilingual text recognition

As per the evaluation in Section 4, it is evident that most of the present applications lack multilingual recognition support. The development of accurate Optical Character Recognition (OCR) engines for regional languages will surely help in making these AI/CV-based apps truly global and inclusive. With recent trends, it is evident that regional language users on digital spaces such as apps and websites have increased exponentially [38]. Designing multilingual interface will not only help in reaching out to larger user bases but it will also accelerate economical and societal growth without any bias.

### 6.6 Indoor navigation assistance

This is one of the futuristic and advanced features presently being developed for these AI/CV-based mobile applications. By leveraging the onboard sensors and camera lens, directional guidance for indoor navigation can be provided to people who have blindness or low vision. Presently only Seeing AI by Microsoft has this experimental feature.

Suggestions and improvements proposed above are purely from the software point of view. Advancements in camera lenses, high-end processors, and other sensors will absolutely transform this paradigm of AI/CV-based applications. Since hardware aspects are out of the scope of this study, no suggestion is proposed on these tracks.

## 7. CONCLUSION AND FUTURE WORK

Mobile applications designed using AI and CV are trending across the globe for their futuristic approach. These solutions are making headlines in the disability community as these are great enablers. This paper has evaluated the current state of AI/CV-based assistive mobile apps for the visually impaired and presented a comprehensive analysis of the existing challenges faced by the target user group. Observations recorded in this paper were both quantitative and qualitative. The evaluation was greatly user-driven and result-oriented, as real beneficiaries (users with visual disabilities) were directly involved. The study also highlights the limitations and drawbacks of these apps in terms of accuracy, performance, privacy concerns, and unfriendly user interfaces. The study shows that visually impaired persons find the accuracy and reliability of AI/CV-based apps as the primary challenges. Other factors hindering their usage are complex design, privacy and safety concerns.

The results presented in this paper demonstrate that despite significant progress in AI and CV-driven knowledge and engineering, there is still great scope for improvement in terms of reliable and performance-oriented application design. This study also emphasises a user-centric approach, greater transparency in data collection, processing and continuous monitoring of app performance to ensure optimal user experiences. By taking a holistic approach to AI and CV-based research, the accuracy, efficiency, reliability and overall usability of these innovative applications can be improved. Technological breakthroughs have always played a pivotal role in societal transformation, and AI-CV will take this change to the next level, especially for visually challenged people.

## ACKNOWLEDGEMENT

This work is supported by the "Knowledge and resource centre for Accessibility in ICT (KAI)", an initiative of the "Ministry of Electronics and Information Technology (MeitY), Government of India" to formulate Indian 'Accessibility Standard for ICT Products and Services'.

## AUTHORS

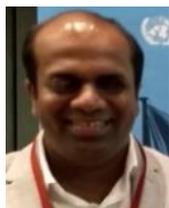
**Saidarshan Bhagat** completed his bachelor's degree in computer application in 2010 and master's in computer application in 2013. He is a certified 'Web Accessibility Specialist' from the International Association of Accessibility Professionals [IAAP] and is [Information Technology Infrastructure Library [ITIL] V3 Foundation certified. He has over 13 years of experience in IT, ICT, development and testing with special expertise in accessibility and assistive technologies. He has worked at ATOS for over a decade before joining C-DAC as a senior technical officer; he is presently working as a principal technical officer. He has several papers to his credit including a conference paper at Melbourne, Australia titled "Evaluation of Accessibility and Accessibility Audit Methods for e-Governance Portals".

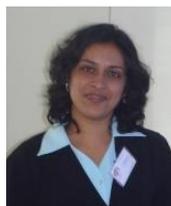
**Dr. Padmaja Joshi** completed her electronics and telecommunication degree in 1992 from COEP Technological University, Pune, her ME (Electronics) from VJTI, Mumbai in 1998 and her Ph.D in software reengineering from the Indian Institute of Technology, Bombay in 2008. She currently works at the Centre for Development of Advanced Computing, Mumbai as a senior director. Her interest areas are software engineering, mobile software engineering, mobile cloud computing, e-governance, blockchain technology, cybersecurity.

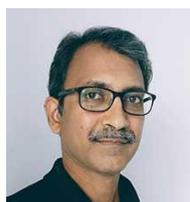
**Avinash Agarwal** completed his Bachelor of Engineering in Electronics & Communication from MNIT Jaipur in 1992 and an M.Tech. (computer science and technology) from the Indian Institute of Technology, Roorkee in 1994. He is an Indian Telecom service officer of the 1992 batch. He has over 29 years of extensive experience in telecommunications, information technology and broadcasting, while serving in various positions in the Government of India. Presently, he is posted as Deputy Director General (Convergence & Broadcasting) at the Telecommunication Engineering Centre, India's Standards Setting Organization. He is Chair of the National Working Group-9, India on "Broadband Cable & TV". He is also Rapporteur Q11/9 and Associate Rapporteur Q3/9 in ITU-T Study Group 9.

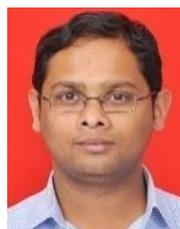
**Shubhanshu Gupta** completed his Bachelor of Technology degree in information technology in 2007. He joined C-DAC in 2008 as a technical officer and is presently serving as a principal technical officer. He has over 15 years of experience in IT solutions with a focus on TV broadcast media accessibility solutions. His interests include ICT and media accessibility solutions and ICT-based assistive technologies.